\documentclass[aps,prb,twocolumn,floatfix,superscriptaddress]{revtex4}

\usepackage{graphicx}
\usepackage{bbm}
\usepackage{amsmath,amssymb}
\newcommand{\be}{\begin{equation}}
\newcommand{\beq}{\begin{equation}}
\newcommand{\ee}{\end{equation}}
\newcommand{\bea}{\begin{eqnarray}}
\newcommand{\eea}{\end{eqnarray}}
\newcommand{\ba}{\begin{array}}
\newcommand{\ea}{\end{array}}

\renewcommand{\vr} {{\bf r}}
\newcommand{\vj} {{\bf j}}
\newcommand{\vs} {{\bf s}}

\begin{document}
\title{Exchange-energy functionals for finite
  two-dimensional systems}
\author{S. Pittalis}
\email[Electronic address:\;]{pittalis@physik.fu-berlin.de}
\affiliation{Institut f{\"u}r Theoretische Physik,
Freie Universit{\"a}t Berlin, Arnimallee 14, D-14195 Berlin, Germany}
\affiliation{European Theoretical Spectroscopy Facility (ETSF)}
\author{E. R{\"a}s{\"a}nen}
\email[Electronic address:\;]{esa@physik.fu-berlin.de}
\affiliation{Institut f{\"u}r Theoretische Physik,
Freie Universit{\"a}t Berlin, Arnimallee 14, D-14195 Berlin, Germany}
\affiliation{European Theoretical Spectroscopy Facility (ETSF)}
\author{N. Helbig}
\affiliation{Unit{\'e} de Physico-Chimie et de Physique des
  Mat{\'e}riaux, Universit{\'e} Catholique de Louvain, B-1348 Louvain-la-Neuve, Belgium}
\affiliation{European Theoretical Spectroscopy Facility (ETSF)}
\author{E. K. U. Gross}
\affiliation{Institut f{\"u}r Theoretische Physik,
Freie Universit{\"a}t Berlin, Arnimallee 14, D-14195 Berlin, Germany}
\affiliation{European Theoretical Spectroscopy Facility (ETSF)}

\date{\today}

\begin{abstract}
Implicit and explicit density functionals for the exchange energy 
in finite two-dimensional systems are developed following
the approach of Becke and Roussel [Phys. Rev. A {\bf 39}, 3761 (1989)].
Excellent agreement for the exchange-hole potentials and 
exchange energies is found when compared with the exact-exchange 
reference data for the two-dimensional uniform electron gas and 
few-electron quantum dots, respectively. Thereby, this work significantly 
improves the availability of approximate density functionals 
for dealing with electrons in quasi-two-dimensional structures, which have
various applications in semiconductor nanotechnology.
\end{abstract}

%\pacs{}

\maketitle

\section{Introduction}

Since the advent of density-functional theory (DFT)~\cite{dft}
much effort went in the development of approximate functionals
for the exchange and correlation energies. Most of this work
focused on three-dimensional (3D) systems, where considerable advances
beyond the commonly used local (spin) density 
approximation (L(S)DA) were achieved by generalized gradient
approximations, orbital functionals, and hybrid functionals.~\cite{functionals}
For two-dimensional (2D) and low-dimensional
systems in general, such efforts have been relatively scarce, partly due to 
the lack of direct applications before the rapid developments 
in semiconductor technology in the early 1980s. Presently,
the variety of relevant low-dimensional 
systems is very large, including, e.g., 
modulated semiconductor layers and surfaces, quantum-Hall 
systems, spintronic devices, and quantum dots (QDs).
The research of electronic, optical, and magnetic properties 
of low-dimensional structures form already a significant 
contribution to condensed-matter and materials physics.

Semiconductor QDs are finite quasi-2D electron systems 
confined typically in GaAs/AlGaAs heterostructures.~\cite{qd}
Due to their controllability in size, shape, and number of confined
electrons, they have an increasing number of potential
applications in quantum-information technology.
Within the DFT approach, QDs are most commonly
treated using the 2D-LSDA exchange functional derived in 1977 by 
Rajagopal and Kimball,~\cite{rajagopal} which is then 
combined with the 2D-LSDA 
correlation parametrized first by Tanatar and Ceperley~\cite{tanatar}
and later, for the complete range of collinear spin polarization, 
by Attaccalite and co-workers.~\cite{attaccalite} 
Despite the relatively good
performance of LSDA with respect to, e.g., quantum Monte
Carlo calculations,~\cite{henri} there is lack of accurate 2D 
density functionals. Moreover, previous studies have shown that
functionals developed for 3D systems perform poorly when applied to 2D 
systems.~\cite{filippi,kim,pollack}
The exact-exchange functional employed 
within the optimized effective potential method, which automatically
conforms to various dimensionalities, seems an appealing
alternative, and it has recently been applied to QDs.~\cite{nicole} 
In that method, however, the development of approximations for the correlation 
energies compatible with exact-exchange energies remains
a complicated problem. 

%A generally accepted alternative procedure is to replace the exact
%expression for the exchange energy functional with an approximation that may be
%more suitable for a corresponding approximation for the correlation energy functional.
%Along these lines, we
%consider the framework  proposed by Becke and Roussel~\cite{becke}, and adapt it
%to 2D systems. We then present an {\em ab initio} derivation of
%new exchange functionals for finite 2D systems, which are natural basis for
%developing corresponding correlation-energy functionals.

In this work we present an {\em ab initio} derivation of 
new exchange functionals for finite 2D systems in the framework
proposed by Becke and Roussel.~\cite{becke}
We calculate the cylindrical average of the exchange hole
for a single-electron wave function of a 2D harmonic
oscillator, and use it as a basis of an averaged exchange
hole for a generic $N$-electron system. By utilizing
the short-range behavior of the exchange hole we are
able to derive one implicit and one explicit density 
functional for the exchange energy.
Both functionals lead to accurate exchange-hole potentials and
exchange energies when compared with the reference data
of the 2D electron gas (2DEG) and few-electron QDs.
The new exchange-energy functionals constitute a natural
basis for developing corresponding correlation-energy functionals.

%This paper is organized as follows. In Sec.~\ref{F}, we 
%illustrate the formalism underlying our approximations
%and examine the short-range behavior of the exchange hole
%for 2D systems. In Sec.~\ref{M} we calculate the cylindrical
%average of our exchange-hole model, and utilize its 
%short-range behavior to derive new exchange-energy functionals.
%In Section~\ref{N}, we show that our new approximations lead to accurate 
%exchange-hole potentials and exchange energies when compared
%with the reference data of the 2D electron gas (2DEG) and 
%few-electron QDs. Brief conclusions are given in Section~\ref{CON}.

\section{Formalism}\label{F}

Within the Kohn-Sham (KS) 
method of spin-DFT,~\cite{BarthHedin:72} the ground 
state energy and spin densities $\rho_{\uparrow}(\vr)$ and $\rho_{\downarrow}(\vr)$ 
of a system of $N=N_{\uparrow}+N_{\downarrow}$ interacting electrons are determined. 
The total energy, which is minimized to obtain the
ground-state energy, is written as a 
functional of the spin densities (in Hartree atomic units)
\begin{eqnarray}
E_{v}[\rho_{\uparrow},\rho_{\downarrow}] & = & T_s[\rho_{\uparrow},\rho_{\downarrow}] + 
E_{\rm H}[\rho] + E_{xc}[\rho_{\uparrow},\rho_{\downarrow}] \nonumber \\
& + & \sum_{\sigma=\uparrow,\downarrow}\int{d\vr} \; v_{\sigma}(\vr)
\rho_{\sigma}(\vr), 
\label{etot}
\end{eqnarray}
where $T_s[\rho_{\uparrow},\rho_{\downarrow}]$ is the KS kinetic energy functional, 
$v_{\sigma}(\vr)$ is the external (local) scalar potential acting 
upon the interacting system, $E_{\rm H}[\rho]$ 
is the classical electrostatic or Hartree energy of the total charge density 
$\rho(\vr)=\rho_{\uparrow}(\vr)+\rho_{\downarrow}(\vr)$, and  
$E_{xc}[\rho_{\uparrow},\rho_{\downarrow}]$ is the exchange-correlation 
energy functional. The latter can be further decomposed
into the exchange and correlation parts as
\be
E_{xc}[\rho_{\uparrow},\rho_{\downarrow}] = E_{x}[\rho_{\uparrow},\rho_{\downarrow}] + E_{c}[\rho_{\uparrow},\rho_{\downarrow}].
\ee
The exchange-energy functional can be expressed as
\begin{equation}
E_x [\rho_{\uparrow},\rho_{\downarrow}] = - \frac{1}{2} \sum_{\sigma=\uparrow,\downarrow} \int d\vr_1 \int d\vr_2 \frac{\rho_{\sigma}(\vr_1)}{|\vr_1-\vr_2|}
h^{\sigma}_{x}(\vr_1,\vr_2),
\label{EX_2}
\end{equation}
where, within the restriction that the noninteracting ground state
is nondegenerate and hence takes the form of a single Slater determinant, 
the exchange-hole (or Fermi-hole) function $h^{\sigma}_{x}$ is given by
\begin{equation}
h^{\sigma}_{x}(\vr_1,\vr_2) =
\frac{|\sum_{k=1}^{N_\sigma}\psi^*_{k,\sigma}(\vr_1)\psi_{k,\sigma}(\vr_2)|^2}
{\rho_{\sigma}(\vr_1)}.
\label{xhole}
\end{equation}
The sum in the numerator is the
one-body spin-density matrix of the Slater determinant constructed from
the KS orbitals, $\psi_{k,\sigma}$. 
The exchange-hole function as defined here is always positive. 
Moreover, integrating this function over $\vr_2$ yields
\begin{equation}
\int d\vr_2 h^{\sigma}_x(\vr_1,\vr_2) = 1.
\label{norm}
\end{equation}
This exact property reflects the fact that around an electron
with spin $\sigma$ at $\vr_1$, other electrons of the same spin
are less likely to be found as a consequence of the Pauli principle.
From Eq.~(\ref{EX_2}) it is clear that to evaluate the exchange energy 
in 2D, we just need to know the {\em cylindrical} average w.r.t. $\vs=\vr_2-\vr_1$
of the exchange-hole around $\vr_1$. Expressing the exchange-hole 
by its Taylor expansion
\begin{equation}
h^{\sigma}_x(\vr_1,\vr_2=\vr_1+\vs) = \exp(\vs\cdot\nabla') h^{\sigma}_x(\vr_1,\vr')|_{\vr'=\vr_1},
\label{taylor}
\end{equation}
the cylindrical average is defined as
\begin{equation}
\bar{h}^\sigma_x(\vr_1,s) = \frac{1}{2\pi}
\int_{0}^{2\pi} d\phi_s \exp(\vs\cdot\nabla') h^\sigma_x(\vr_1,\vr')|_{\vr'=\vr_1}.
\label{avxhole}
\end{equation}
The short-range behavior with respect to $s$ is then obtained as
\begin{eqnarray}
\bar{h}^{\sigma}_x(\vr_1,s) & = & \rho_{\sigma}(\vr_1)
+ \frac{s^2}{4} \nabla'^2 h^{\sigma}_x(\vr_1,\vr')|_{\vr'=\vr_1} +
\ldots \nonumber \\
& = & \rho_{\sigma}(\vr_1) + s^2 C^{\sigma}_{x}(\vr_1) + \ldots ,
\label{taylor2}
\end{eqnarray}
where $C^{\sigma}_x$ is the so-called local curvature of the exchange
hole around the given reference point $\vr_1$. This function can be expressed
as~\cite{becke,becke2,Dobson:93}
\begin{equation}
C^{\sigma}_x = \frac{1}{4}\left[ \nabla^2 \rho_{\sigma} -2\tau_\sigma
+ \frac{1}{2}\frac{\left( \nabla \rho_\sigma \right)^2}{\rho_\sigma}
+ 2 \frac{\vj^2_{p,\sigma}}{\rho_\sigma} \right],
\label{C}
\end{equation}
where
\begin{equation}
\tau_\sigma=\sum_{k=1}^{N_\sigma} |\nabla\psi_{k,\sigma}|^2
\end{equation}
is (twice) the spin-dependent kinetic-energy density, and 
\begin{equation}
\vj_{p,\sigma}=\frac{1}{2i}\sum_{k=1}^{N_\sigma} \left[
  \psi^*_{k,\sigma} \left(\nabla \psi_{k,\sigma}\right) - \left(\nabla \psi^*_{k,\sigma}\right) \psi_{k,\sigma} \right]
\end{equation}
is the spin-dependent paramagnetic current density.
Both $\tau_\sigma$ and
$\vj_{p,\sigma}$
depend explicitly on the KS orbitals. Thus the expression in Eq.~(\ref{C})
has an implicit dependence on the spin-densities $\rho_\sigma$.
Now, once we can provide an approximation for 
$\bar{h}^{\sigma}_x(\vr_1,s)$ satisfying the normalization condition
of Eq.~(\ref{norm}), we can compute the exchange-hole potential
\begin{equation}
U^{\sigma}_x(\vr_1)= - 2 \pi \int_{0}^{\infty}  ds \,\bar{h}^{\sigma}_x(\vr_1,s),
\label{xenergydensity}
\end{equation}
and finally the exchange energy
\begin{equation}
E_x[\rho_{\uparrow},\rho_{\downarrow}] =  \frac{1}{2} \sum_{\sigma} \int d\vr_1 \rho_{\sigma}(\vr_1) 
U^{\sigma}_x(\vr_1).
\label{xenergy}
\end{equation}

\section{Exchange-energy functional}\label{M}

\subsection{Model}

As the basis of our exchange model we
consider the ground state single-electron wave function of a 2D harmonic oscillator 
\begin{equation}
\psi_\sigma(\vr) = \frac{\alpha}{\sqrt{\pi}} \exp \left[-\frac{\alpha^2 r^2}{2}\right],
\end{equation}
which is normalized for each $\alpha \ne 0$.
We point out that
the harmonic (parabolic) approximation for the confining potential is the most
common choice when modeling QDs fabricated 
in semiconductor heterostructures.~\cite{qd}
The exact exchange-hole function (\ref{xhole}) for the single-particle
case becomes
$h^{\sigma}_{x}(\vr_1,\vr_2)=\psi^*_{\sigma}(\vr_2)\psi_{\sigma}(\vr_2) 
= \rho_\sigma(\vr_2)$. Setting $\vr_1=\vr$ and $\vr_2=\vr+\vs$, 
we calculate the cylindrical average as
\begin{eqnarray}\label{SEC}
\bar{h}^\sigma_x(\vr,s) & = & 
 \frac{1}{2\pi} \int_{0}^{2\pi} d\phi_s \,\rho_\sigma(\vr+\vs) \nonumber \\
& = & 
\frac{\alpha^2}{2\pi^2} \exp \left[-\alpha^2
\left(r^2+s^2\right)\right] \times \nonumber \\
& \times & 
\int_{0}^{2\pi} d\phi_s \exp \left[-2\alpha^2rs\cos(\phi)\right]
\nonumber \\
& = &
\frac{\alpha^2}{\pi} \exp \left[-\alpha^2 \left( r^2 + s^2 \right) \right]
I_0(2\alpha^2rs),
\end{eqnarray}
where $I_0(x)$ is the zeroth order modified Bessel function of the first kind
(note that $I_0(0)=1$). Performing the integral
in Eq.~(\ref{xenergydensity}) yields
\begin{equation}
U^{\sigma}_x(r)=-|\alpha|\sqrt{\pi}\exp\left[-{\alpha^2 r^2/2}\right] I_0(\alpha^2 r^2/2)
\label{model}
\end{equation}
for the exchange-hole potential. Since the modified Bessel
function has the limit property
$I_0(x)\rightarrow \exp(x)/\sqrt{2\pi x}$, 
when $x \rightarrow +\infty$ with
$x \in {\mathbb R}$, we immediately
find $\lim_{r\rightarrow\infty} U^{\sigma}_x(r)=-1/r$.
%Furthermore, the resulting total energy is
%self-interaction free by construction.

\subsection{Implicit density functional}
\label{implicit}

At this point, we adopt the strategy of Becke and Roussel~\cite{becke} and
elevate expression (\ref{SEC}) as a general model
for the averaged exchange hole of a generic $N$-electron 2D system.
In order to locally reproduce the
short-range behavior of the exchange hole, 
we replace $\alpha^2$ and $r^2$ by functions of $r$,
respectively, i.e., $\alpha^2\rightarrow a(r)$ and $r^2\rightarrow b(r)$.
Now we can rewrite Eq.~(\ref{SEC}) as
\begin{equation}
\bar{h}^{\sigma}_x(a,b;s) = \frac{a}{\pi} \exp \left[-a \left( b + s^2 \right) \right] I_0(2a\sqrt{b}s).
\label{h}
\end{equation}
This model satisfies, through its original
definitions, $\bar{h}^{\sigma}_x(a,b;s)\ge0$ and the unit normalization
condition of Eq.~(\ref{norm}). From the second-order term in the
Taylor expansion in Eq.~(\ref{taylor2}) we obtain
\begin{equation}
\left(y-1\right)\exp(y) = \frac{C^{\sigma}_{x}}{\pi\rho^2_{\sigma}},
\label{relation}
\end{equation}
where $y:=ab$. The zeroth-order term gives
\be
a = \pi\rho_{\sigma}\exp{(y)},
\label{a}
\ee
and hence we get
\be
b = \frac{y}{\pi\rho_{\sigma}}\exp{(-y)}.
\label{b}
\ee
As the result, Eqs.~(\ref{relation})-(\ref{b}) determine,
together with Eqs.~(\ref{h}), (\ref{xenergydensity}) 
and (\ref{xenergy}), an implicit density functional.
This completes the derivation of our first
approximation to the exchange-energy functional.

\subsection{Explicit density functional}
\label{explicit}

We now show that one can also derive an explicit density
functional. For that purpose,
we consider the 2DEG where the derivatives in 
Eq.~(\ref{C}) are zero, i.e.,
$\nabla^2\rho_{\sigma}=0$ and $\nabla\rho_{\sigma}=0$, and
we take the known expression for the kinetic-energy
density of the 2DEG,\cite{Zyl}
$\tau_{\sigma}=2\pi\rho^2_{\sigma}+\vj^2_{p,\sigma}/\rho_\sigma$.
Note that the current-dependent term in Eq.~(\ref{C}) cancels
the current-contribution to the kinetic energy, leading
to the expected result that the exchange energy of a uniform
electron gas does not depend on the current 
density.~\footnote{A uniform current in a homogeneous electron
gas can be viewed as being generated by a transformation to a 
moving Galilei frame. Since the electron-electron Coulomb interaction
is the same in any Galilei frame, the exchange energy cannot 
depend on the Galilei frame and, hence, must be current-independent.}
These simplifications lead to $y=0$, $b=0$, and $a=\pi\rho_{\sigma}$,
so that the averaged exchange-hole function becomes
\begin{equation}
\bar{h}^{\sigma}_{x}(s)=\rho_{\sigma}\exp\left[-\pi\rho_{\sigma}s^2 \right],
\label{final_xhole}
\end{equation}
and the exchange-hole potential is now given by
$U^{\sigma}_{x}[\rho_{\sigma}]=-\pi\rho_{\sigma}^{1/2}$, which is
an explicit functional of the spin-density.
Defining the 2D density parameter $r_s=1/\sqrt{\pi\rho}$, where 
$\rho=\rho_{\uparrow}+\rho_{\downarrow}$ is the total density, and
the polarization
$\xi=(\rho_{\uparrow}-\rho_{\downarrow})/\rho$, the total exchange
energy per particle becomes
\begin{equation}
\epsilon_{x}[r_s,\xi]=-\frac{\sqrt{\pi}}{4\sqrt{2}\,r_s}\left[(1+\xi)^{3/2}+(1-\xi)^{3/2}\right].
\label{final_xenergy_N}
\end{equation}
This expression can be compared with the exact 2DEG
result~\cite{rajagopal} widely applied to finite systems 
in terms of the LSDA,
\begin{equation}
\epsilon^{\rm 2DEG}_{x}[r_s,\xi]=-\frac{2\sqrt{2}}{3\pi r_s}\left[(1+\xi)^{3/2}+(1-\xi)^{3/2}\right].
\label{2DEG_xenergy}
\end{equation}
%Note that we use Hartrees as energy units.
Interestingly, the only difference between these two
expressions for $\epsilon_x$ is in the prefactor, which is 
%which is $\sim 0.3133$ and $\sim
%0.3001$ in our functional and in the 2DEG (LSDA) form, 
%respectively, 
about 
%implying a difference of 
$\sim 4.4\%$ smaller in the latter. Concluding, Eq.~(\ref{final_xenergy_N})
can be used as an explicit density functional in the LSDA fashion
when applied to inhomogeneous systems.

\section{Numerical results}\label{N}

\subsection{Two-dimensional electron gas}

\begin{figure}
\includegraphics[width=0.95\columnwidth]{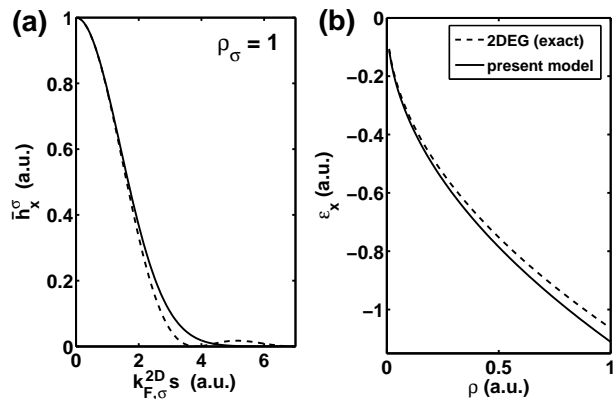}
\caption{Exchange hole (a) and the exchange
energy per particle (b) for two-dimensional electron gas ($\xi=0$)
in the exact expression (dashed line) and
in the present model (solid line).}
\label{2deg}
\end{figure}

Besides the exchange energies, it is interesting to compare
the averaged exchange hole given in Eq.~(\ref{final_xhole}) 
with the exact exchange hole of the 2DEG. 
%The latter quantity,
%well known in 3D~\cite{slater}, has been rarely considered in 
%2D~\cite{ghosh}. 
Following the derivation of Gori-Giorgi and co-workers~\cite{gori} for 
the pair-distribution functions of the 2DEG, we find 
\begin{equation}
\bar{h}^{\rm 2DEG}_{x,\sigma}(s)=4\rho_{\sigma}\frac{J^2_1(k^{\rm 2D}_{F,\sigma} s)}{(k^{\rm 2D}_{F,\sigma} s)^2},
\label{2DEG_xhole}
\end{equation}
where $k^{\rm 2D}_{F,\sigma}=(4\pi\rho_{\sigma})^{1/2}$ is the
Fermi momentum (for spin $\sigma$) 
in 2D, and $J_1$ is the ordinary Bessel function 
of the first order.
%Note that the 3DEG expression is of a similar form, but in that
%case the prefactor is nine, the Bessel function is spherical, and
%the Fermi momentum is $k^{\rm 3D}_F=(6\pi^2\rho_{\sigma})^{1/3}$ (see
%Ref.~\cite{slater}). 
In Fig.~\ref{2deg}(a) 
we compare the exchange holes with a fixed spin-density 
$\rho_{\sigma}=1$ between our model (solid line) 
and the exact 2DEG result (dashed line) and find
a good qualitative agreement.
Figure~\ref{2deg}(b) demonstrates the differences 
in the exchange energies (per particle) given in Eqs.~(\ref{final_xenergy_N}) and 
(\ref{2DEG_xenergy}) as a function of the total density (for $\xi=0$).

\subsection{Few-electron quantum dots}

\begin{figure}
\includegraphics[width=0.9\columnwidth]{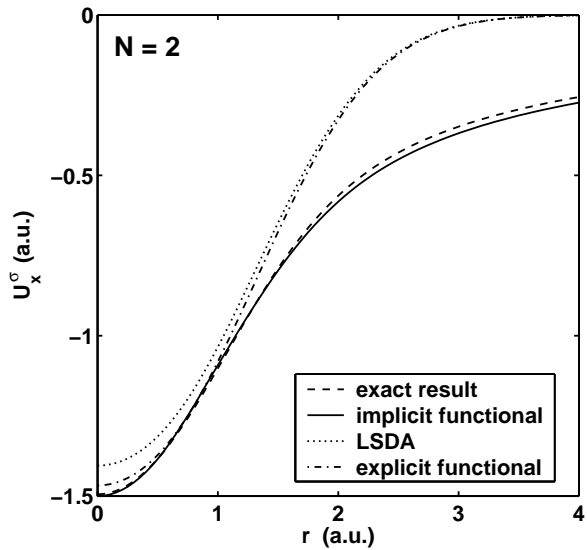}
\caption{Exchange-hole potential of
a two-electron quantum dot calculated exactly (dashed line)
and using the implicit density functional 
of Sec.~\ref{implicit} (solid line), standard LSDA
(dotted line), and the explicit density functional 
of Sec.~\ref{explicit} (dash-dotted line).
The analytic density and orbitals
are used as input for the model and LSDA.}
\label{n2}
\end{figure}

Next, we consider the smallest nontrivial QD consisting of two 
electrons. The Hamiltonian is given by
\begin{equation}
H=\sum_{j=1}^2\left(-\frac{1}{2}\nabla^2_j + \frac{1}{2}\omega^2_0 r_j^2\right)+\frac{1}{|\vr_2-\vr_1|},
\end{equation}
where we set the strength of the harmonic confinement to $\omega_0=1$.
In this case the ground-state (singlet) wave function is known
analytically,~\cite{taut} and the total density 
%\begin{equation}
%\Psi(\vr_1,\vr_2)=\frac{\exp\left[-\frac{1}{2}(r_1^2+r_2^2)\right]}{\pi\sqrt{\sqrt{2\pi}+3}}
%\end{equation}
%and the analytic ground-state density 
can be expressed as~\cite{wensauerthesis}
\begin{eqnarray}
\rho(r) & = & \frac{4}{\pi(\sqrt{2\pi}+3)}\Big\{e^{-r^2}(1+r^2/2)+
\frac{1}{2}\sqrt{\pi}e^{-3r^2/2} \times \nonumber \\ 
& \times & \Big[I_0(r^2/2)+r^2 I_0(r^2/2)+r^2 I_1(r^2/2)\Big]\Big\}.
\end{eqnarray}
The exact (spin) exchange-hole potential is simply
\begin{eqnarray}
U^{x,\sigma}_{N=2}(\vr)=-\int d\vr'\frac{\rho_{\sigma}(\vr')}{|\vr-\vr'|}.
\end{eqnarray}
In Fig.~\ref{n2} 
we compare the exact exchange-hole potential (dashed line)
to the result of our implicit density functional
(solid line) and find an excellent agreement. 
We note that in the range $0.75 \lesssim r\lesssim 0.85$
the parameter $y$ is not solvable from 
Eq.~(\ref{relation}). Therefore, we set $y$ to zero 
in this region. This corresponds to the 2DEG result discussed above, 
i.e., we perform a well-valid LSDA-type approximation
in this small regime.
The dash-dotted line shows the result of our explicit
density functional which is identical to the implicit 
functional in the 2DEG limit.
This density functional has the wrong asymptotic behavior, 
but it is considerably closer to the exact result than 
that of the standard LSDA of Eq.~(\ref{2DEG_xenergy}) (dotted line).

\begin{figure}
\includegraphics[width=0.9\columnwidth]{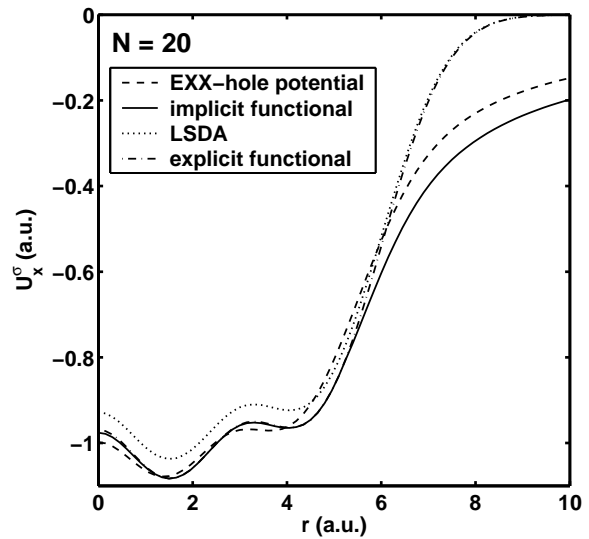}
\caption{ 
Similar to Fig.~\ref{n2} but for 20 electrons. 
The exact-exchange result was calculated 
using the KLI approximation. 
Self-consistent (correlation included) density and orbitals from
the standard LSDA are used as input for our functionals and LSDA
exchange.}
\label{n20}
\end{figure}

In Fig.~\ref{n20} 
we show the exchange-hole potentials for a 
20-electron QD with $\omega_0=0.42168$, corresponding
to a typical confinement of 5 meV in the effective-mass 
approximation when modeling QDs in GaAs.~\cite{qd}
The exact-exchange (EXX) result (dashed line) is
calculated in the Krieger-Li-Iafrate (KLI)
approach~\cite{KLI} within spin-DFT
implemented in the {\tt octopus} real-space code.~\cite{octopus} 
Again, we find a very good agreement between the EXX
and the results of our implicit density
functional (solid line). Also, the explicit
functional (dash-dotted line) reproduces the EXX curve
at small $r$ by a good accuracy, whereas the LSDA
curve (dotted line) lies considerably above the
EXX result in this regime. However, 
as seen in Table~\ref{table}
\begin{table}%[width=\columnwidth]
\caption{\label{table}
Exchange energies for different number of electrons
calculated using the exact exchange,
the implicit density functional of Sec.~\ref{implicit}, standard LSDA, and the 
the explicit density functional of Sec.~\ref{explicit}. 
For $N>2$ the EXX result was calculated within the KLI approximation,
and as input for our functionals and LSDA exchange we used the
self-consistent density and orbitals from the standard LSDA.
}
\begin{tabular*}{\columnwidth}{@{\extracolsep{\fill}} c c c c c}
\hline
\hline
N & EXX & Implicit functional & LSDA & Explicit functional \\
\hline
2  & -1.0839 & -1.0836 & -0.983 & -1.026 \\
6  & -2.229  & -2.284  & -2.130 & -2.223 \\
12 & -4.890  & -5.059  & -4.763 & -4.972 \\
20 & -8.781  & -9.124  & -8.632 & -9.012 \\
\hline
\hline
\end{tabular*}
\end{table}
the LSDA exchange energy is rather accurate.
This is due to the fact that the 
difference in the EXX and LSDA curves at small
$r$ is compensated at $5\lesssim r\lesssim 6$, where the
LSDA curve lies below the EXX one.
In smaller QDs instead, the table shows that 
our exchange-energy functionals are superior to 
the LSDA also in terms of the exchange energies, i.e.,
not only in terms of the exchange-hole potentials.

\section{Conclusions}\label{CON}

To conclude, we have provided new exchange functionals
for finite two-dimensional systems that, until now,
have lacked accurate functionals beyond the 
local density approximations. We have presented
{\em ab initio} derivation of both implicit and explicit 
exchange functionals that significantly improve the
exchange-hole potentials and exchange energies 
of few-electron two-dimensional quantum dots.
Our results also suggest that 
accurate approximations for the correlation energy functional
can be developed analogously.

\begin{acknowledgments}
We thank S. Kurth, A. Castro, and R. Hammerling for useful discussions
and help in the numerical calculations. 
This work was supported by the Deutsche 
Forschungsgemeinschaft, the EU's Sixth Framework
Programme through the Nanoquanta Network of
Excellence (No. NMP4-CT-2004-500198), the Academy of
Finland, and the Finnish Academy of Science and Letters
through the Viljo,
Yrj{\"o} and Kalle V{\"a}is{\"a}l{\"a} Foundation.
\end{acknowledgments}

\end{document}